\documentclass[%
superscriptaddress,
amsmath,
amssymb,
aps,
twocolumn,
longbibliography
]{revtex4-1}

\usepackage{array}
\usepackage{wrapfig}
\usepackage{graphicx}
\usepackage{dcolumn}
\usepackage{bm}
\usepackage{amsmath}
\usepackage{amssymb}
\usepackage{graphics}
\usepackage{epsfig}
\usepackage{natbib}
\usepackage{verbatim} 
\usepackage{color}

\usepackage{tikz} 
\usepackage{etoolbox} 

\newcommand{\muSR}{$\mu$SR }
\newcommand{\Rb}{RbEuFe$_\mathrm{4}$As$_\mathrm{4}$ }
\newcommand{\Tc}{$T_{c}$ }
\newcommand{\TC}{$T_{m}$ }
\newcommand{\muSRp}{$\mu$SR}
\newcommand{\Rbp}{RbEuFe$_\mathrm{4}$As$_\mathrm{4}$}
\newcommand{\Tcp}{$T_{c}$}
\newcommand{\TCp}{$T_{m}$}

\newrobustcmd*{\mysquare}[1]{\tikz{\filldraw[draw=#1,fill=#1] (0,0) rectangle (0.2cm,0.2cm);}}
\newrobustcmd*{\mycircle}[1]{\tikz{\filldraw[draw=#1,fill=#1] (0,0) circle [radius=0.1cm];}}
\newrobustcmd*{\mytriangle}[1]{\tikz{\filldraw[draw=#1,fill=#1] (0,0) -- (0.282cm,0) -- (0.141cm,0.244cm);}}
\newrobustcmd*{\mytdiamond}[1]{\tikz{\filldraw[draw=#1,fill=#1] (0,0.141cm) -- (0.141cm,0) -- (0.283cm,0.141cm) -- (0.141cm,0.283cm);}}
\newrobustcmd*{\mytstar}[1]{\tikz{\filldraw[draw=#1,fill=#1] (0,0.158cm) -- (0.218cm,0) -- (0.135cm,0.256cm) -- (0.051cm,0) -- (0.269cm,0.158cm);}}
\definecolor{aqua}{rgb}{0.0, 1.0, 1.0}

\begin{document}

\title{Muon spin rotation measurements on RbEuFe$_\mathrm{4}$As$_\mathrm{4}$ under pressure}

\author{S.~Holenstein}
\email{stefan.holenstein@psi.ch}
\affiliation{Laboratory for Muon Spin Spectroscopy, Paul Scherrer Institute,
CH-5232 Villigen PSI, Switzerland}
\affiliation{Physik-Institut der Universit\"at Z\"urich,
Winterthurerstrasse 190, CH-8057 Z\"urich, Switzerland}
\author{B.~Fischer}
\affiliation{Department Chemie, Ludwig-Maximilians-Universit\"at M\"unchen,
Butenandtstrasse 5-13 (D), 81377 M\"unchen, Germany}
\author{Y.~Liu}
\affiliation{Department of Physics, Zhejiang University, Hangzhou 310027, China}
\author{N.~Barbero}
\affiliation{Laboratorium f\"ur Festk\"orperphysik, ETH Z\"urich, CH-8093
Zurich, Switzerland}
\author{G.~Simutis}
\affiliation{Laboratory for Muon Spin Spectroscopy, Paul Scherrer Institute,
CH-5232 Villigen PSI, Switzerland}
\author{Z.~Shermadini}
\affiliation{Laboratory for Muon Spin Spectroscopy, Paul Scherrer Institute,
CH-5232 Villigen PSI, Switzerland}
\author{M.~Elender}
\affiliation{Laboratory for Muon Spin Spectroscopy, Paul Scherrer Institute,
CH-5232 Villigen PSI, Switzerland}
\author{P.~K.~Biswas}
\affiliation{ISIS Pulsed Neutron and Muon Source, STFC Rutherford Appleton Laboratory,
Harwell Campus, Didcot, Oxfordshire OX11 0QX, United Kingdom}
\author{R.~Khasanov}
\affiliation{Laboratory for Muon Spin Spectroscopy, Paul Scherrer
Institute, CH-5232 Villigen PSI, Switzerland}
\author{A.~Amato}
\affiliation{Laboratory for Muon Spin Spectroscopy, Paul Scherrer
Institute, CH-5232 Villigen PSI, Switzerland}
\author{T.~Shiroka}
\affiliation{Laboratory for Muon Spin Spectroscopy, Paul Scherrer
Institute, CH-5232 Villigen PSI, Switzerland}
\affiliation{Laboratorium f\"ur Festk\"orperphysik, ETH Z\"urich, CH-8093
Zurich, Switzerland}
\author{H.-H.~Klauss}
\affiliation{Institute of Solid State and Materials Physics, TU Dresden, DE-01069
Dresden, Germany}
\author{E.~Morenzoni}
\affiliation{Laboratory for Muon Spin Spectroscopy, Paul Scherrer Institute,
CH-5232 Villigen PSI, Switzerland}
\affiliation{Physik-Institut der Universit\"at Z\"urich,
Winterthurerstrasse 190, CH-8057 Z\"urich, Switzerland}
\author{G.-H.~Cao}
\affiliation{Department of Physics, Zhejiang University, Hangzhou 310027, China}
\author{D.~Johrendt}
\affiliation{Department Chemie, Ludwig-Maximilians-Universit\"at M\"unchen,
Butenandtstrasse 5-13 (D), 81377 M\"unchen, Germany}
\author{H.~Luetkens}
\email{hubertus.luetkens@psi.ch}
\affiliation{Laboratory for Muon Spin Spectroscopy, Paul Scherrer Institute,
CH-5232 Villigen PSI, Switzerland}

\begin{abstract}

We report muon spin rotation and magnetization measurements on the magnetic superconductor RbEuFe$_{4}$As$_{4}$ under hydrostatic pressures up to 3.8\,GPa. At ambient pressure, RbEuFe$_\mathrm{4}$As$_\mathrm{4}$ exhibits a superconducting transition at $T_{c} \approx$ 36.5\,K and a magnetic transition at $T_{m} \approx$ 15\,K below which the magnetic and the superconducting order coexist. With increasing pressure, $T_{c}$ decreases while $T_{m}$ and the ordered Eu magnetic moment increase. In contrast to iron-based superconductors with ordering Fe moments, the size of the ordered Eu moment is not proportional to $T_{m}$. The muon spin rotation signal is dominated by the magnetic response impeding the determination of the superconducting properties.

\end{abstract}

\maketitle

\section{Introduction}

In 2009, it was found that isovalent P substitution on the As site in EuFe$_2$(As$_{1-x}$P$_x$)$_2$ suppresses the spin density wave order of the Fe moments and changes the antiferromagnetic order of the Eu moments to ferromagnetic order which coexists with superconductivity for a small substitution range $x$ \cite{Ren2009a,Nowik2011}.
There has been a vivid debate how superconductivity and ferromagnetism can coexist in this so-called 122 system.
One plausible theory states that the superconducting pairing and the ferromagnetic coupling of the Eu moments through the Ruderman-Kittel-Kasuya-Yosida (RKKY) interaction involve different Fe-3\textit{d} orbitals.
While it is mainly the $d_{x^2-y^2}$ and $d_{z^2}$ orbitals, which provide the RKKY coupling, the superconducting pairing is dominated by the $d_{yz}$ and $d_{zx}$ orbitals \cite{Cao2011}. In recent years, the intrinsically hole-doped iron-based superconductor \Rbp, an intergrowth of EuFe$_{2}$As$_2$ and RbFe$_{2}$As$_2$, has attracted a significant amount of attention due to its comparably high superconducting transition temperature \Tc $\approx 36.5$\,K and the coexistence of superconducting and magnetic order below \TC $\approx 15$\,K \cite{Liu2016d,Kawashima2016a}.
The anion heights, i.e. the heights of the As above the Fe plane, are close to the empirical optimum of 1.38\,\AA\,\cite{Mizuguchi2010,Okabe2010} to achieve highest \Tc \cite{Liu2016d,Bao2018}.
It was shown that the in-plane ferromagnetic order in this compound is associated solely with the Eu magnetic moments that are aligned perpendicularly to the crystallographic $\textit{c}$-axis \cite{Albedah2018b}. The three dimensional magnetic structure is still under debate, however, with some studies arguing in favor of ferromagnetic order \cite{Liu2016d,Albedah2018b} while a recent study claims a helical antiferromagnetic structure \cite{Iida2019}. The not so common coexistence of magnetism with superconductivity calls for microscopic investigations of \Rbp.

In this work we present a combination of local-probe muon spin rotation and relaxation (\muSRp) measurements and magnetization measurements on \Rb under hydrostatic pressures up to 3.8\,GPa. We find that \Tc decreases with pressure while \TC increases, in agreement with data from literature \cite{Jackson2018,Xiang2019}. In addition, our local-probe \muSR measurements show that the ordered magnetic moment increases by about 4\% at 2.4\,GPa, while \TC increases by 24\%. We do not find any signature of a significant coupling between the superconducting and the magnetic order.

\section{Experimental methods}

Polycrystalline \Rb was synthesized via a solid-state reaction method \cite{Liu2016d} and characterized using powder X-ray diffraction (PXRD). \muSR measurements were performed at the Swiss Muon Source (S$\mu$S) using the General Purpose Surface-Muon (GPS) \cite{Amato2017} and the General Purpose Decay-Channel (GPD) \cite{Khasanov2016d} spectrometers. The data were analyzed with the free software package \textsc{musrfit} \cite{Suter2012}. Magnetization measurements were performed using a commercial vibrating sample magnetometer (VSM) and a superconducting quantum interference device (SQUID) magnetometer. Hydrostatic pressure for the \muSR  measurements was applied using a double-wall piston cell made from MP35N alloy \cite{Khasanov2016d} with Daphne 7373 oil \cite{Yokogawa2007} as a pressure transmitting medium. A CuBe anvil-type cell with CuBe gaskets, self-aligning ZrO$_{2}$ anvils, and Daphne 7575 oil \cite{Murata2016} as a pressure transmitting medium was used for magnetization measurements. Pressures were determined by either In (\muSRp) or Pb (SQUID) manometers \cite{Schilling1981c}.

\section{Results}

\begin{figure}[tb]
\centering{
\includegraphics[width=1\columnwidth]{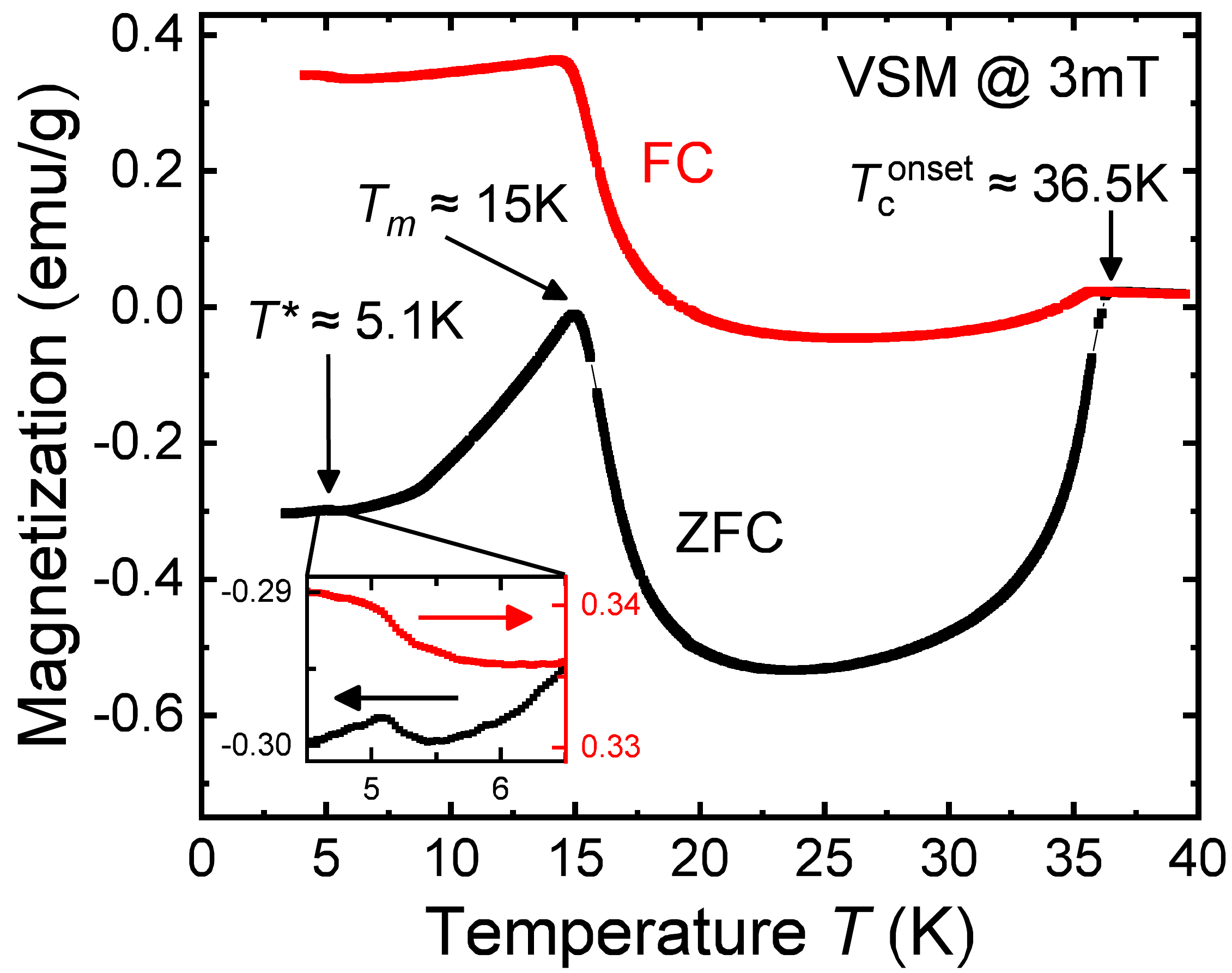}
 \caption{Magnetization of \Rb at ambient pressure measured by vibrating sample magnetometry in 3\,mT [cooled in zero field (ZFC) or in applied field (FC)] showing the superconducting transition at \Tc and the magnetic transition at \TCp. Inset: Magnetization in the temperature range around the impurity related anomaly at $T$* for ZFC [black (dark), left axis] and FC [red (light), right axis] measurements.
 }\label{Fig:VSM}
}
\end{figure}

VSM measurements shown in Fig. \ref{Fig:VSM} confirm a superconducting transition temperature \Tc $\approx 36.5$\,K and a magnetic transition temperature \TC $\approx 15$\,K. Further, there is a small anomaly at $T$* $\approx 5.1$\,K (inset Fig. \ref{Fig:VSM}). This anomaly was previously observed by magnetization and heat capacity measurements \cite{Liu2016d} and was later realized to be likely due to very small amounts of Eu$_{3}$O$_{4}$ impurities \cite{Liu2017} which order antiferromagnetically at $T_{N} \approx 5$\,K \cite{Holmes1966}.

\muSR measurements, which require a comparably large amount of sample when performed under pressure, were carried out on 1.65\,g of \Rb with 6.4\% RbFe$_\mathrm{2}$As$_\mathrm{2}$ and 6.7\% EuFe$_\mathrm{2}$As$_\mathrm{2}$ impurities. A small amount of Eu$_{3}$O$_{4}$ impurity below the detection limit of the characterizing PXRD measurements was presumably present too, given the anomaly at $T$* mentioned before. Representative zero-field (ZF) \muSR spectra recorded with no external magnetic field applied are shown in Fig. \ref{Fig:spectra} for temperatures above and below \TC $\approx 15$\,K. Below the magnetic transition temperature, spontaneous muon spin precession can be observed due to the static long range magnetic order. The data were analysed using two different models for the temperatures above and below \TCp. Above \TCp, a simple phenomenological model was applied:

\begin{figure}[tb]
\centering{
\includegraphics[width=1\columnwidth]{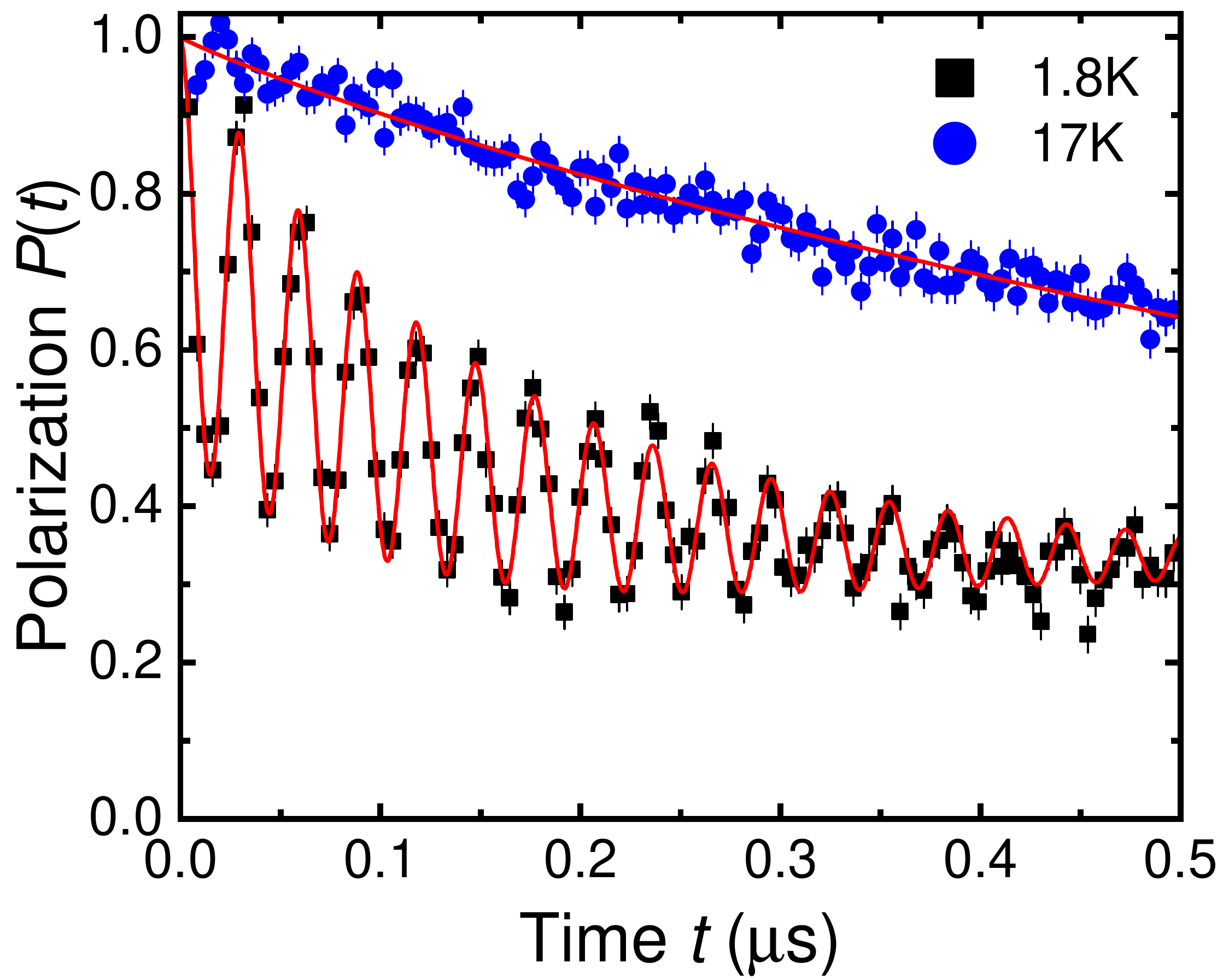}
 \caption{Representative zero-field \muSR spectra measured at ambient pressure above and below the magnetic transition temperature \TC $\approx 15$\,K. The oscillations at low temperature indicate static long range magnetic order. The red lines are fits using the models introduced in Eqs. (\ref{Eq:ZF-fitHT}) and (\ref{Eq:ZF-fitLT}).
 }\label{Fig:spectra}
}
\end{figure}

\begin{equation}\label{Eq:ZF-fitHT}
  P_\mathrm{HT}(t)=(1-f_{1})\exp[-(\lambda_\mathrm{HT} t)^\beta]+\,f_{1},
\end{equation}

where $\lambda_\mathrm{HT}$ is the relaxation rate and $\beta$ is a stretching exponent. $f_{1}$ is a small nonrelaxing tail fraction due to the already present static magnetic order of the EuFe$_\mathrm{2}$As$_\mathrm{2}$ impurity \cite{Raffius1993,Ren2008,Guguchia2013}. The data below \TC were modelled by:

\begin{equation}\label{Eq:ZF-fitLT}
  \begin{split}
  P_\mathrm{LT}(t)=&\,\frac{2}{3}[f_\mathrm{osc}\cos(\gamma_{\mu}B_\mathrm{int}t)\exp(\lambda_{T}t)\\
  &\,+(1-f_\mathrm{osc})\exp(\lambda_\mathrm{no}t)]+\frac{1}{3}\exp(\lambda_{L}t)\,,
  \end{split}
\end{equation}

where the 2/3 (transverse) and 1/3 (longitudinal) components reflect the powder average of the internal fields with respect to the initial muon spin direction in a the polycrystalline sample. The transverse part consists of an oscillating fraction $f_\mathrm{osc} \approx 0.4$ and a nonoscillating fraction $f_\mathrm{no}=1-f_\mathrm{osc} \approx 0.6$. $\lambda_{T}$, $\lambda_\mathrm{no}$, and $\lambda_{L}$ are the corresponding relaxation rates. The oscillation frequency is given by $\gamma_{\mu}B_\mathrm{int}$, where $\gamma_{\mu}=2\pi\times135.5$\,MHzT$^{-1}$ is the muon's gyromagnetic ratio and $B_\mathrm{int}$ is the magnetic field at the muon stopping site. The latter is proportional to the ordered magnetic moment and therefore a measure of the magnetic order parameter. The oscillating signal from the few percent of magnetically ordered EuFe$_\mathrm{2}$As$_\mathrm{2}$ and Eu$_{3}$O$_{4}$ impurities was not included in the analysis as it was too small to be resolved. Possible contributions from the impurity phases are absorbed by the last two terms of Eq. \ref{Eq:ZF-fitLT}. An influence on the determination of $B_\mathrm{int}$ is very unlikely due to the significantly higher internal fields in EuFe$_\mathrm{2}$As$_\mathrm{2}$ \cite{Guguchia2013,Tran2018} and the very small amount of Eu$_{3}$O$_{4}$. In the case of measurements under hydrostatic pressure, the signal from muons stopping in the pressure cell was treated in analogy to Ref. \cite{Khasanov2016d}.

\begin{figure}[tb]
\centering{
\includegraphics[width=1.0\columnwidth]{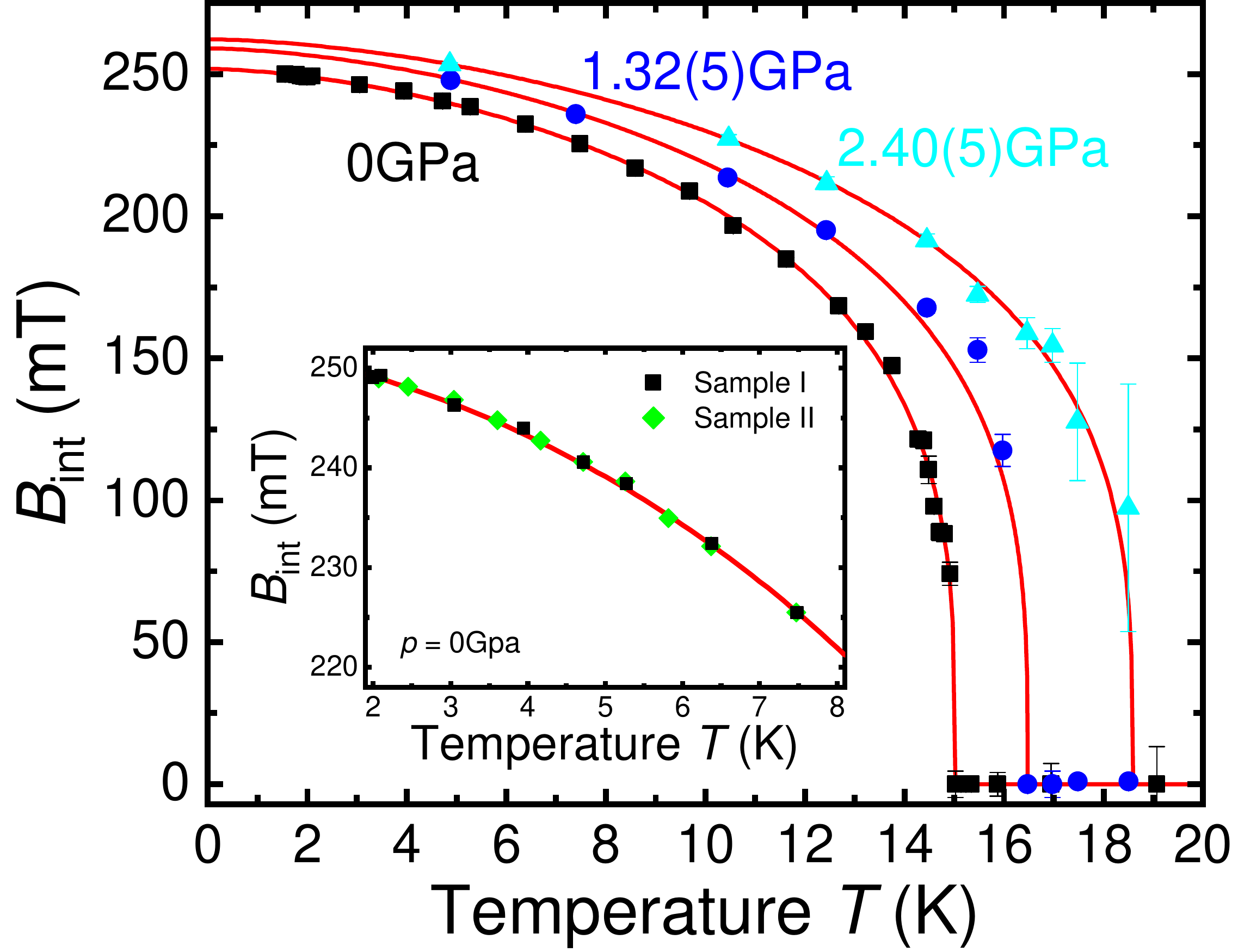}
 \caption{Temperature dependence of the internal magnetic field $B_\mathrm{int}$ at the muon stopping site for different pressures. $B_\mathrm{int}$ is proportional to the ordered magnetic moment and therefore a measure of the magnetic order parameter. The red lines are fits using the model described in Eq. (\ref{Eq:Bint}). Inset: No anomaly in $B_\mathrm{int}$ is observed around $T$*.
 }\label{Fig:Bint}
}
\end{figure}

\begin{figure}[tb]
\centering{
\includegraphics[width=1.0\columnwidth]{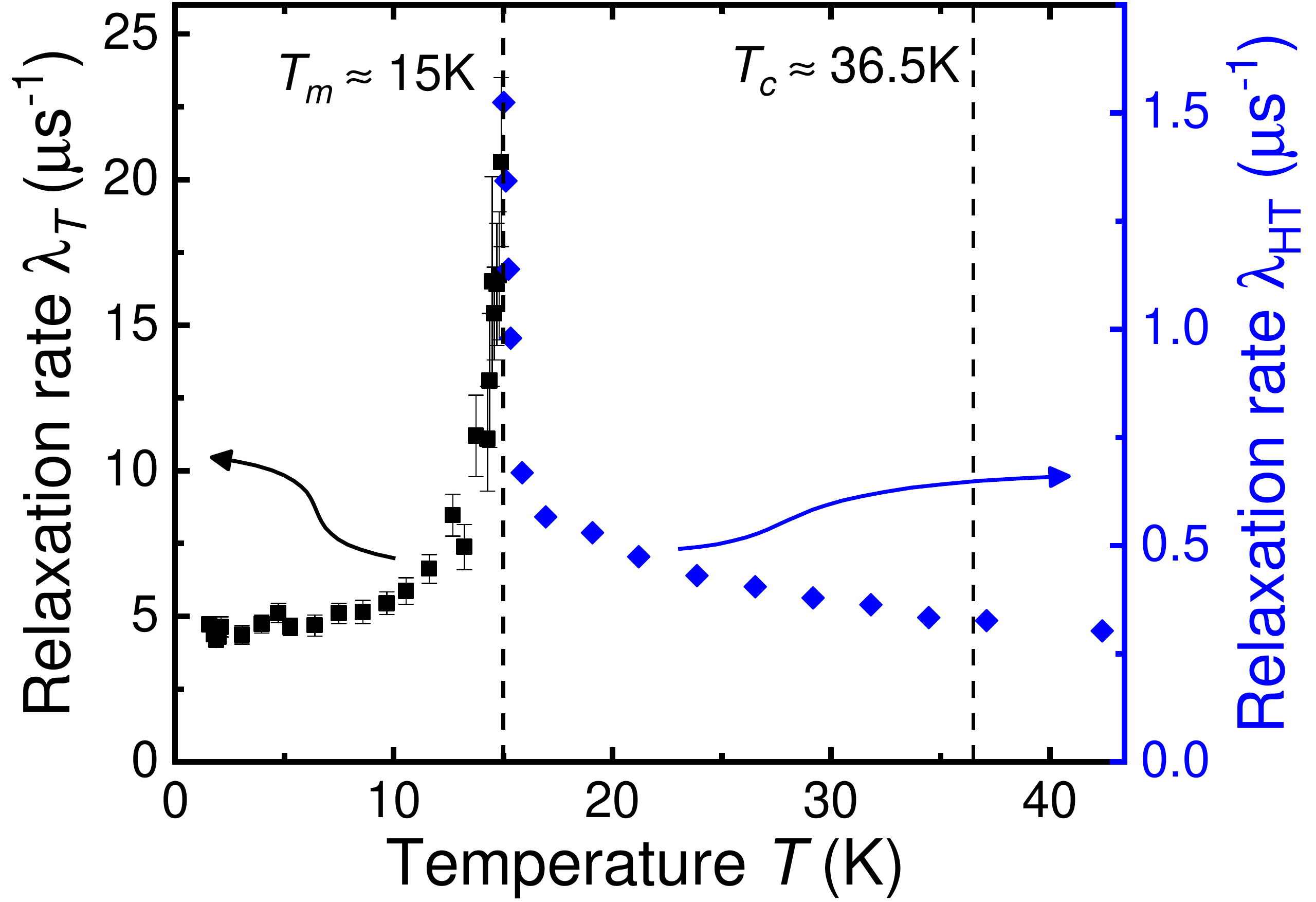}
 \caption{Temperature dependence of the muon spin relaxation rates obtained by fitting the zero-field \muSR data measured at ambient pressure to Eqs. (\ref{Eq:ZF-fitHT}) and (\ref{Eq:ZF-fitLT}).
 }\label{Fig:lambda}
}
\end{figure}

Fig. \ref{Fig:Bint} shows the temperature dependence of $B_\mathrm{int}$ for different pressures. The red lines are fits using the phenomenological function \cite{Tran2018}:

\begin{equation}\label{Eq:Bint}
  B_\mathrm{int} = B_\mathrm{int,0}(1-(T/T_{m})^{\alpha})^{\gamma},
\end{equation}

where $B_\mathrm{int,0}$ is the field at zero temperature. $\alpha = 1.63(4)$ and $\gamma = 0.29(1)$ were determined from the ambient pressure data and fixed for the fit of the pressure data. The magnetic transition temperature increases monotonically with pressure in agreement with literature data \cite{Jackson2018,Xiang2019}. Simultaneously, our ZF \muSR measurements show that the ordered magnetic moment is enhanced. At 2.4\,GPa, the increase amounts to about 24\% for the transition temperature \TCp, but only about 4\% for the magnetic moment.
The inset of Fig. \ref{Fig:Bint} includes data measured on a second batch of \Rb and focuses on $B_\mathrm{int}$ in the temperature region around the feature at $T$* $\approx 5.1$\,K observed by VSM (inset Fig. \ref{Fig:VSM}). The lack of an anomaly in the temperature dependencies of the internal field $B_\mathrm{int}$ and the transverse relaxation rate $\lambda_{T}$ (Fig. \ref{Fig:lambda}) rules out a change in the magnetic structure like a spin reorientation and therefore supports the notion of impurities as a cause for this feature.

The fraction $f_\mathrm{no}=1-f_\mathrm{osc}$ in Eq. (\ref{Eq:ZF-fitLT}) describes those parts of the sample that are magnetic but too disordered to exhibit coherent muon spin oscillations (correlation length smaller than approximately 10 lattice constants \cite{Yaouanc2011}). The corresponding relaxation rate $\lambda_\mathrm{no}$ sharply increases to about 9\,$\mu$s$^{-1}$ within the first Kelvin below \TC and stays roughly constant at this value for lower temperatures (not shown). The longitudinal relaxation rate $\lambda_{L}$ (not shown), which can be nonzero only for dynamic systems, drops quickly to zero below \TCp, indicating that the whole volume of the sample, including $f_\mathrm{no}$, is static below \TCp. In the temperature region between \TC and \Tcp, magnetic fluctuations lead to a sizable and temperature dependent relaxation rate $\lambda_\mathrm{HT}$ (Fig. \ref{Fig:lambda}) which renders an investigation of the superconducting properties of \Rb by the means of \muSR unfeasible.

\begin{figure}[b]
\centering{
\includegraphics[width=1.0\columnwidth]{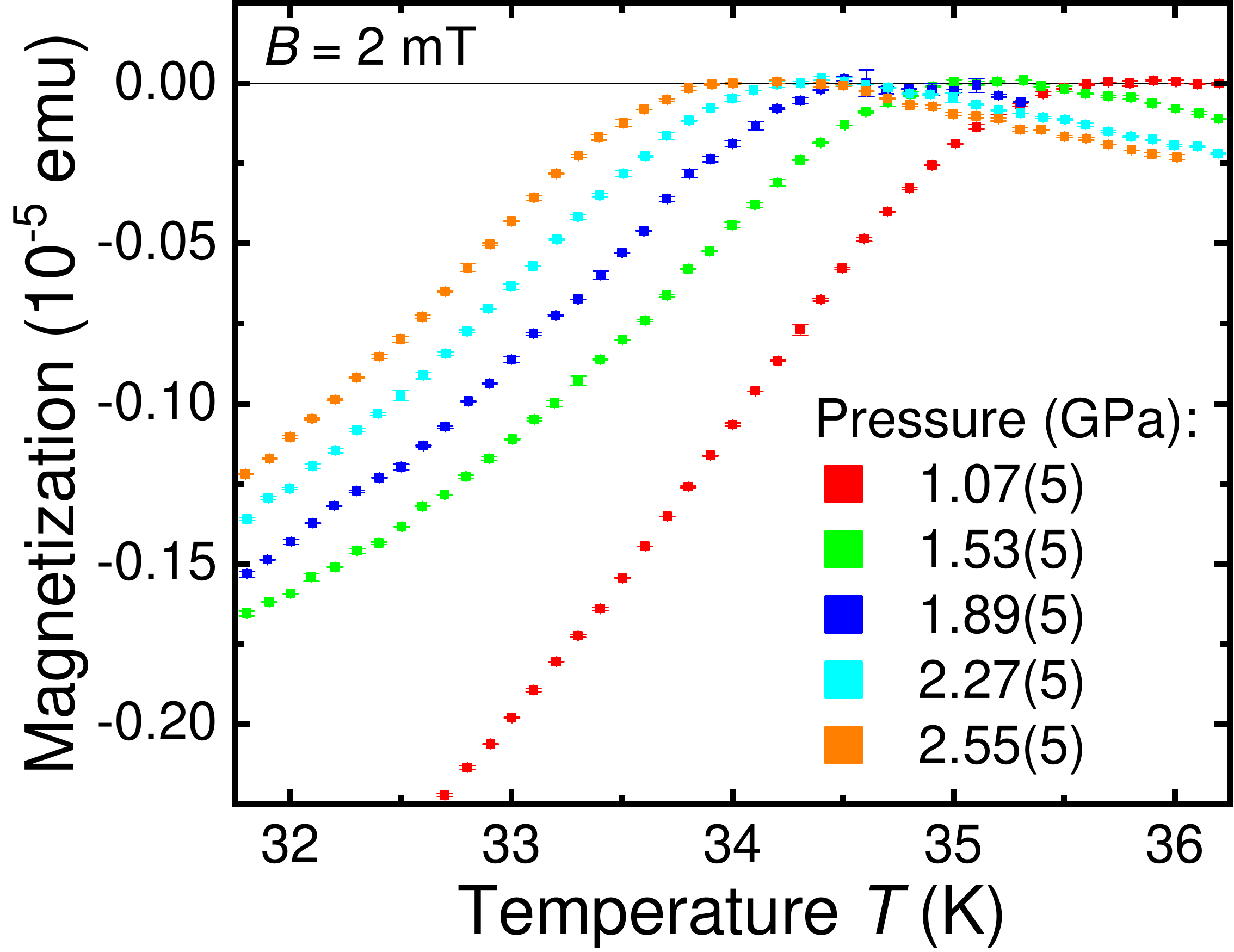}
 \caption{Magnetization of a small grain of \Rb as a function of temperature for a representative series of pressures measured by SQUID magnetometery in a field of 2\,mT. For all measurements, the sample was cooled in zero field. The pressure cell background was subtracted and the data were shifted vertically to overlap around \Tcp.
 }\label{Fig:SQUID}
}
\end{figure}

\begin{figure}[tb]
\centering{
\includegraphics[width=1.0\columnwidth]{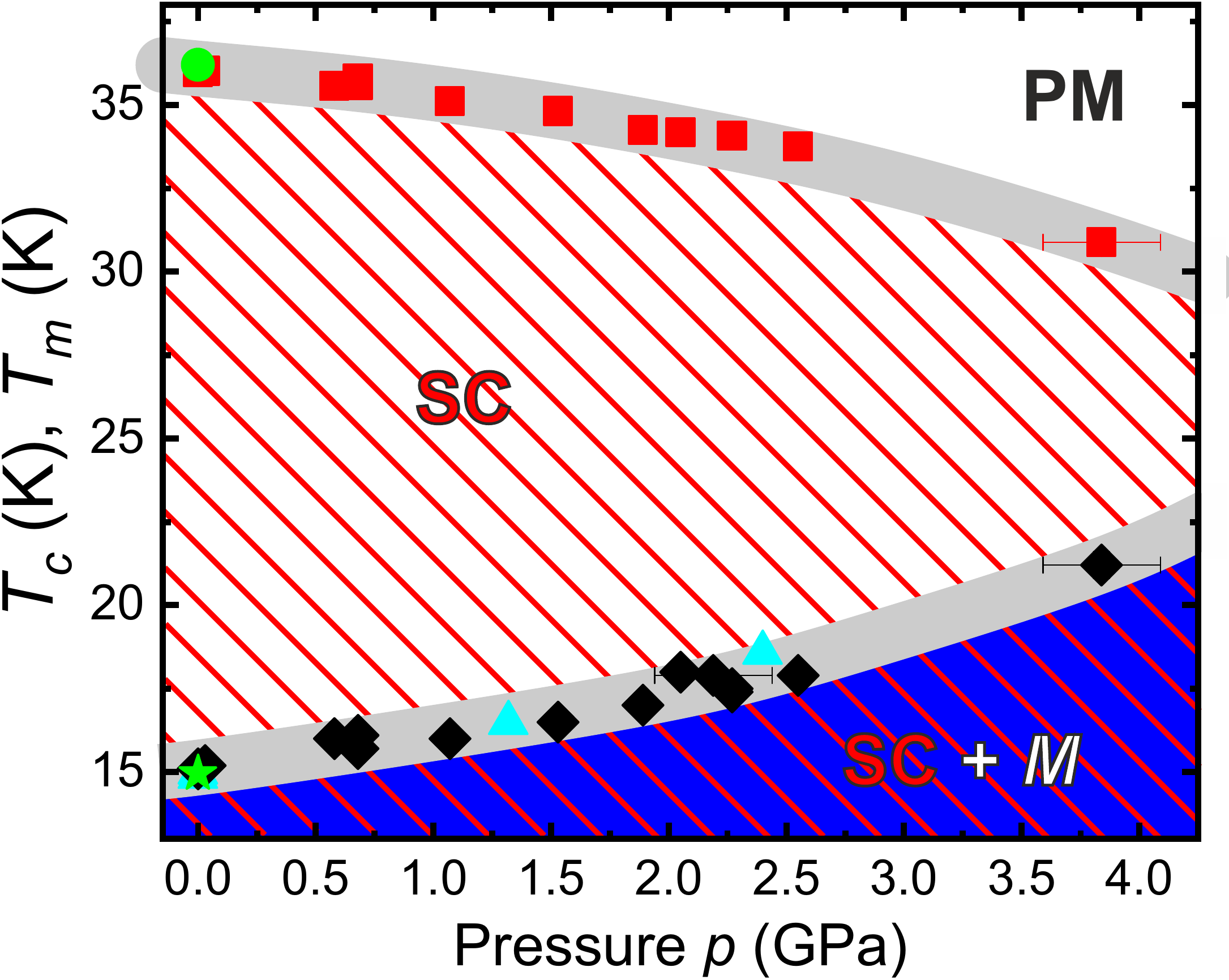}
 \caption{Temperature-pressure phase diagram of \Rbp. Superconducting transition temperatures \Tc were measured by SQUID magnetometry (\mysquare{red}) and vibrating sample magnetometry (\mycircle{green}). Magnetic transition temperatures \TC were measured by SQUID magnetometry (\mytdiamond{black}), vibrating sample magnetometry (\mytstar{green}), and zero-field muon spin rotation measurements (\mytriangle{aqua}). With increasing hydrostatic pressure, \Tc decreases while \TC increases. PM and SC denote the paramagnetic and the superconducting phase respectively. SC+\textit{M} denotes the region of coexisting superconducting and magnetic order. The gray shaded areas are guides to the eye.
 }\label{Fig:overview}
}
\end{figure}

We therefore employed SQUID magnetometry to determine the pressure dependence of \Tc and to further investigate the magnetic transition. Magnetization data for the temperature range around \Tc are shown in Fig. \ref{Fig:SQUID} for a representative series of pressures. The cell background was subtracted and the data were shifted vertically to overlap around the transition. \Tc was determined by the intersection of two linear approximations of the data above and below the transition. The same method was used to determine \TC (not shown). The results are presented in the temperature-pressure phase diagram shown in Fig. \ref{Fig:overview}, together with the transition temperatures obtained by ZF \muSR and VSM.

\section{Discussion}

Under hydrostatic pressure, the superconducting transition temperature \Tc of \Rb decreases monotonically, while the magnetic transition temperature \TC increases (Fig. \ref{Fig:overview}), in agreement with literature data \cite{Jackson2018,Xiang2019}. Additionally, our ZF \muSR measurements reveal sizable magnetic fluctuations already above \TC and show that the ordered magnetic moment also increases under hydrostatic pressure. At 2.4\,GPa, \TC is enhanced by about 24\% compared to ambient pressure, but the ordered magnetic moment increases by only about 4\%. This nonproportional relation sets \Rb apart from magnetic members of the 122 and other families of iron-based superconductors where a proportional scaling of the two quantities was found \cite{Uemura2009,Cheung2018}. In these systems, the magnetic order is usually associated with Fe moments, whereas in \Rb magnetism is due to the ordering of Eu moments \cite{Albedah2018b}, which might explain the different scalings.

Despite the seemingly antagonistic behavior of the superconducting and the magnetic state in \Rb our findings imply that there is no significant coupling between the two orders in agreement with Refs. \cite{Kawashima2018,Smylie2018,Iida2019,Xiang2019}. The rate of decrease in \Tc is comparable to nonmagnetic CaKFe$_\mathrm{4}$As$_\mathrm{4}$ under pressure \cite{Kaluarachchi2017}. Therefore, the suppression of superconductivity is unlikely due to the enhanced magnetic order but rather caused by the pressure induced changes of the lattice parameters \cite{Jackson2018} which likely drive the anions away from the optimal height value. Further, substitution studies show that the superconducting and the magnetic order can be suppressed independent of each other \cite{Liu2017,Kawashima2018}.
This is in agreement with various mechanisms proposed to explain the coupling among the Eu moments, namely the so-called \textit{d}-\textit{f} \cite{Kasuya1970} interaction or As-Eu-As superexchange interactions as proposed in Ref. \cite{Liu2017} or an RKKY interaction that involves different Fe-3\textit{d} orbitals than the superconducting pairing as proposed in Refs. \cite{Cao2011,Liu2016d}.
Subtle effects like small influences of the onset of magnetic order on the vortex lattice like described in Refs. \cite{Stolyarov2018,Vlasko-Vlasov2019} are not detectable by \muSR however due to the dominance of the magnetic signal.

Neither the magnetic order parameter measured via the zero-field muon spin precession frequency nor the muon spin relaxation rates, which reflect the field distribution at the muon stopping sites, exhibit anomalies in the low temperature region (inset of Fig. \ref{Fig:Bint}, Fig. \ref{Fig:lambda}). The anomaly at $T$* $\approx 5.1$\,K in magnetization measurements (inset Fig. \ref{Fig:VSM}) is therefore clearly not related to the magnetic order in \Rbp. This supports the attribution to small Eu$_{3}$O$_{4}$ impurities mentioned in Ref. \cite{Liu2017}. Eu$_{3}$O$_{4}$ represents the most likely impurity, not only due to the antiferromagnetic order below $T_{N} \approx 5$\,K \cite{Holmes1966}, but also due to the fact that the $T$* anomaly is reported for the magnetic superconductor CsEuFe$_\mathrm{4}$As$_\mathrm{4}$ as well \cite{Liu2016c}, but not for the Eu free members of the 1144 family. In contrast, a connection to the EuFe$_\mathrm{2}$As$_\mathrm{2}$ or the RbFe$_\mathrm{2}$As$_\mathrm{2}$ impurity phase seems unlikely since no features are reported around $T$* in the literature for these compounds \cite{Raffius1993,Ren2008,Guguchia2013,Bukowski2010,Moroni2019}.

\section{Conclusion}

In conclusion, we have shown that the superconducting order in \Rb is suppressed by the application of hydrostatic pressure while the magnetic order is enhanced. The relation between the magnetic transition temperature \TC and the size of the ordered magnetic moment is not proportional, setting \Rb apart from magnetic members of the 122 and other families of iron-based superconductors, where a proportional scaling of the two quantities was found \cite{Uemura2009,Cheung2018}. No significant coupling between the magnetic and the superconducting order was found in agreement with earlier reports \cite{Kawashima2018,Smylie2018,Iida2019,Xiang2019}.


\begin{acknowledgments}
This work is partially based on experiments performed at the Swiss Muon Source S$\mu$S, Paul Scherrer Institute, Villigen, Switzerland. We gratefully acknowledge the financial support of S.H. and N.B. by the Swiss National Science Foundation (SNF-Grant Nos. 200021-159736 and 200021-169455).
\end{acknowledgments}



%


\end{document}